\documentclass[12pt]{article}    % Specifies the document style.
\usepackage{amsfonts,latexsym}

\begin{document}

\title{Solvable models of Bose-Einstein condensates: a new algebraic
Bethe ansatz scheme
}
\author{Huan-Qiang Zhou\footnote{email: hqz@maths.uq.edu.au},
Jon Links\footnote{email: jrl@maths.uq.edu.au}, 
Mark D. Gould and Ross H.  McKenzie \\ 
Centre for Mathematical Physics,\\  
School of Physical Sciences,\\ 
The University of Queensland,
4072,\\  Australia 
}
%\date{April 15, 1993}

\def\Z{{\mathbb Z}} 

\maketitle

%\vspace{10pt}

\begin{abstract}
A new algebraic Bethe ansatz scheme is proposed to diagonalise 
classes of integrable models relevant to the description of 
Bose-Einstein condensation in  
dilute alkali
gases. This is achieved by introducing the notion of  $\Z$-graded
representations of the Yang-Baxter algebra.
\end{abstract}

%************************** Text Begins here ******************************

%  Greek letters

\def\a{\alpha}
\def\b{\beta}
\def\d{\dagger}
\def\e{\epsilon}
\def\g{\gamma}
\def\K{\kappa}
\def\l{\lambda}
\def\o{\omega}
\def\t{\theta}
\def\s{\sigma}
\def\D{\Delta}
\def\L{\Lambda}
\def\ap{\approx} 
\def\A{{\mathcal {A}}}
\def\N{{\mathcal {N}}}
\def\M{{\mathcal {M}}}
\def\I{{\mathcal {I}}}
% Shorthands for \begin{equation} and the like

\def\beq{\begin{equation}}
\def\eeq{\end{equation}}
\def\bea{\begin{eqnarray}}
\def\eea{\end{eqnarray}}
\def\ba{\begin{array}}
\def\ea{\end{array}}
\def\no{\nonumber}
\def\le{\langle}
\def\re{\rangle}
\def\lt{\left}
\def\rt{\right}
\def\o{\omega}
\def\d{\dagger}
\def\nn{\nonumber} 
\def\Z{{\mathbb Z}}

\newcommand{\reff}[1]{eq.~(\ref{#1})}

%\newpage
\vskip.3in

\section{Introduction}

Theoretical studies     
into the behaviour of Bose-Einstein condensates (BECs) continue at a
prolific rate, motivated by
the experimental successes of producing 
condensates of atomic alkali gases \cite{and,anglin} and superpositions 
of atomic-molecular alkali gases 
\cite{zoller,donley}. Many of the theoretical 
results to date have been obtained
through use of the Gross-Pitaevskii mean-field theory and
generalisations (e.g., see \cite{mcww,pit99,leggett,cusack,vardi}).
However, such mean-field theory 
approaches have limited applicability in regions of the
parameter space where quantum fluctuations dominate. In these cases, only
an exact treatment of the model will give a reliable description of the
physics. 

In our recent work we have shown that one model describing Josephson
tunneling between two coupled BECs \cite{lz,zlmx}, and
another that models coherent superpositions of atomic and molecular
BECs \cite{zlm},  
can both in fact be solved exactly in
the framework of the algebraic Bethe ansatz. Our intention here is to
develop a new mathematical approach which allows us to 
extend this method to establish that very general classes of
Hamiltonians for BECs admit exact solutions. These classes of
Hamiltonians cannot be solved in the usual form of the algebraic Bethe
ansatz.  

In this Letter, three classes of solvable models relevant to Bose-Einstein
condensates of dilute alkali gases are determined. This is achieved by
formulating a new scheme for the algebraic Bethe ansatz by introducing
the notion of $\Z$-graded representations of the Yang-Baxter algebra. 
The first model we will present 
is a six-parameter generalisation of the canonical Josephson Hamiltonian
\cite{leggett} (which can also be considered as a two site Bose-Hubbard model) 
describing a tunnel-coupled pair of trapped Bose-Einstein condensates.
The Hamiltonian is also applicable to model solid state Josephson
junctions and coupled Cooper pair boxes \cite{mss}.
The second model we present  has  six free parameters and the third
one has ten parameters. They both describe coherent coupling between atomic
and diatomic molecular BEC's with additional 
interactions such as $S$-wave scattering  between the atoms, between the
molecules, and between atoms and molecules. Such effects were not
included in \cite{vardi,zlm} but are important for a quantitative
description of experiments \cite{kh}. 
Finally, we formulate the Slavnov formula for the
scalar products between a Bethe eigenstate and an arbitrary Bethe vector,
which facilitates the exact computation of form factors and
correlations functions analogous to the results of \cite{lz}. 

\section{Bethe ansatz for 
$\Z$-graded representations of the Yang-Baxter algebra}

The main  ingredient in the study of exactly solvable quantum systems  
through the algebraic Bethe ansatz \cite{kib,faddeev}
is the Yang-Baxter equation   
\beq
R _{12} (u-v)  R _{13} (u)  R _{23} (v) =
R _{23} (v)  R _{13}(u)  R _{12} (u-v). 
\label{ybe} \eeq
Here $R_{jk}(u)$ denotes the matrix in  ${\rm End\,}(
V \otimes V\otimes V)$ acting non-trivially on the
$j$-th and $k$-th spaces and as the identity on the remaining space.
The $R$-matrix solution may be viewed as the structural constants for the
Yang-Baxter algebra, denoted  $\A$, generated by the monodromy matrix $T(u)$
\beq
R_{12}(u-v) T_1(u) T_2(v)=
T_2(v) T_1(u)R_{12}(u-v). \label{YBA}
\eeq
The simplest case is that for  
the $sl(2)$ invariant $R$-matrix,
which will be the subject of our study, given by 
\beq
R(u) = \left ( \begin {array} {cccc}
1&0&0&0\\
0&b(u)&c(u)&0\\
0&c(u)&b(u)&0\\
0&0&0&1\\
\end {array} \right ),
\label{rm} \eeq
with the rational functions $b(u)=u/(u+\eta)$ and 
$c(u)=\eta/(u+\eta)$.

Setting
\beq
T(u) = \left ( \begin {array} {cc}
A(u)&B(u)\\
C(u)&D(u)
\end {array} \right ),\label{mono}
\eeq
it follows from the defining relations (\ref{YBA}) that  
\bea
& & [A(u),\, A(v)] =
    [D(u), \,D(v)] = 0, \no\\
& & [B(u), \,B(v)] =
    [C(u), \,C(v)] = 0, \no\\
& & A(u)C(v) =
\frac {u-v+\eta}{u-v} C(v) A(u)-\frac {\eta}{u-v}
C(u) A(v), \no\\
& & D (u) C(v) =
\frac {u-v-\eta}{u-v} C(v) D(u) +\frac {\eta}{u-v}
C(u) D(v). \label{rels}   
\eea
Note that there are many more relations satisfied by the the
generators of the Yang-Baxter
algebra. However, those given above are the only ones needed for the
algebraic Bethe ansatz procedure which we investigate  below. 
For convenience, we extend $\A$ by a unit element $I$ which we will
represent by the identity matrix in any representation. 

We also introduce an auxiliary operator $Z$, called the {\it grading
operator}, which satisfies the relations 
\beq  [Z,\,X(u)]=p\{X(u)\}.X(u), \label{grading} \eeq  
where $X=A,\,B,\,C$ or $D$ and $p\{A(u)\}=p\{D(u)\}=0,\,p\{B(u)\}=1$ and 
$p\{C(u)\}=-1$.  
We call $p\{X(u)\}\,\in\,\Z$
the {\it gradation} of $X(u)$ and extend the gradation operation
to the entire algebra
by  the requirement 
$$p\{\theta.\phi\}=p\{\theta\}+p\{\phi\}~~~~\forall \,\theta,\,\phi\,\in 
\A . $$ 
This definition for the grading operator is consistent with the defining
relations of $\A$ governed by (\ref{YBA}). 

\def\Z{{\mathbb Z} }

Let us now define a new class of representations of the Yang-Baxter
algebra which we call ${\mathbb Z}$-graded 
representations.  We say that   a vector space $V$, equipped 
with an endomorphism $z$, 
is a $\Z$-graded vector space, denoted $(V,z)$,  
if it admits a
decomposition into subspaces  
$$V=\bigoplus_{k=-\infty}^\infty V_k$$ 
such that 
$$zV_k=k.V_k, ~~~~~~k\,\in\, \Z. $$ 
Note that some of the $V_k$ may be trivial subspaces. 
Formally, the grading operator can be used to define the following
projection operators 
\beq P_k=\prod_{\stackrel{j=-\infty}{j\neq k}}^\infty \frac{(z-jI)}{(k-j)}
\label{proj} \eeq  
such that 
$$P_kP_l=\delta_{kl} P_l,~~~~P_kV_j=\delta_{kj}V_k. $$ 
We say that a $\Z$-graded vector space 
$$V'=\bigoplus_{k=-\infty}^\infty V'_k$$ 
is {\it equivalent} to $V$ if 
for some $j\,\in\,\Z$ there exist a vector space isomorphism between 
$V'_k$ and $V_{j+k}$ for all $k$. 
This terminology is motivated by the fact that for a given $(V,z)$ one
can always generate another $\Z$-graded space $(V',z')$ through 
the mappings 
$V'_k\rightarrow V_{j+k},\,z'\rightarrow z-jI$ for any $j\,\in\,\Z$. 

For a given $\Z$-graded $V$ we say that  
$\pi :\,\A\rightarrow  {\rm End}\, V$ provides a $\Z$-graded
representation of $\A$ if $\pi(Z)=z$ and the relations
(\ref{YBA},\,\ref{grading})
are preserved. In such a case we can write 
$$\pi(X(u))=\sum_{j=-\infty}^\infty X(u,j)$$ 
where the matrices 
$X(u,j)$ satisfy 
$$X(u,j)V_k=0 ~~~~~{\rm for}\,\, j\neq k. $$   
More specifically, this means that for $\left|\psi_k\right>\,\in\,V_k$
we have 
$$\pi(X(u)Y(v))\left|\psi_k\right>
=X(u,k+p\{Y(u)\})Y(v,k)\left|\psi_k\right>. $$ 
In view of the equivalence of $\Z$-graded vector spaces defined above,
there can also exist equivalent representations. We can define   
a representation $\pi'$ equivalent to $\pi$  
by specifying   some $k\,\in\,\Z$ such that 
$$\pi'(Z)= \pi(Z-kI)  $$   
and for 
$$\pi'(X(u))=\sum_{j=-\infty}^\infty X'(u,j)$$
the matrices $X'(u,j)$ are defined by     
$$X'(u,j)= X(u,j+k)~~~~\forall \,j\,\in\,\Z. $$

For any $\Z$-graded  representation it follows from (\ref{rels}) that the
following hold:
\bea
 [A(u, j), A(v, j)] &=&
    [D(u, j), D(v, j)] =0, \no\\
 B(u,j) B(v, j-1)& = &
    B(v,j) B(u, j-1),  \no\\
    C(u,j) C(v,j+1) &=&
    C(v,j) C(u,j+1), \no\\
 A(u, j)C(v, j+1) &=&
\frac {u-v+\eta}{u-v} C(v,j+1) A(u,j+1)\no\\
&&-\frac {\eta}{u-v}
C(u,j+1) A(v,j+1), \no\\
 D (u,j) C(v,j+1)& =&
\frac {u-v-\eta}{u-v} C(v,j+1) D(u,j+1)\no\\
&&+\frac {\eta}{u-v}
C(u,j+1) D(v,j+1),
\eea

From the defining relations (\ref{YBA}) the transfer matrix defined by
$\tau(u) = A(u)+D(u)$ commutes for different 
values of the spectral parameter $u$; 
viz. $$[\tau(u),\tau(v)]=0.$$ Moreover, we may express the representation
$\pi(\tau(u))$ of the transfer matrix as 
$$\pi(\tau(u))=\sum_{j=-\infty}^\infty \tau(u,j) $$ such that 
$$\tau(u,j)V_k=0~~~~~~{\rm for}\,\, j\neq k$$ 
and 
$$[\tau(u,j),\,\tau(v,k)]=0~~~~~\forall\,j,\,k. $$ 
Since $p\{\tau(u)\}=0$, the diagonalisation of $\pi(\tau(u))$ 
is thus reduced to the diagonalisation
of each of the matrices $\tau(u,j)$ on the $\Z$-graded component $V_j$,
where we have 
$$[\tau(u,j),\,\tau(v,j)]=0. $$ 
We may restrict our attention to the
case of $\tau(u,0)$, as each $\tau(u,j)$ is equivalent to some
$\tau'(u,0)$ through the use of equivalent representations as introduced
earlier. 

In order to formulate the algebraic Bethe ansatz solution for this class
of representations, we assume the existence of a pseudovacuum vector 
$|\chi \rangle \in V_k$ such that 
\bea 
A(u,k) |\chi \rangle &=&   \alpha(u,k)|\chi \rangle \nn \\
B(u,k) |\chi \rangle &=& 0 \nn \\
C(u,k) |\chi \rangle &\neq & 0 \nn \\
D(u,k) |\chi \rangle &=&   \delta(u,k)|\chi \rangle. \nn 
\eea 
The above implies that $|\chi \rangle$ is a maximal weight vector
with respect to $Z$. Without loss of generality we can choose $k=M$,   
again due to the equivalence of representations discussed earlier,  
and look for Bethe states  defined by  
\bea  
\Psi (v_1,\cdots, v_M) &\equiv& \Psi(\{v_i\}) \nn \\
&=& C(v_1,1)C(v_2,2)\cdots C(v_M,M)|\chi \rangle
\label{estates}.
\eea  
It is easy to check that this Bethe state is symmetric with respect
to the variables $v_i$, a feature which plays a crucial role below. 
Acting $A(u,0)$ 
and $D(u,0)$ on the Bethe state
we have
\bea
&&A(u,0) \Psi(\{v_i\})=
\a (u,M) \prod ^M_{i=1} \frac {u-v_i+\eta}{u-v_i}  
\Psi(\{v_i\})\no\\
&&~~~~~~~+\sum^M_{i=1} \M_i (u,\{v_j\})
\Psi(v_1,\cdots,v_{i-1},u,v_{i+1},\cdots,v_M),\no\\
&& D(u,0) \Psi(v_1,\cdots,v_M)=
\delta (u,M) \prod ^M_{i=1} \frac {u-v_i-\eta}{u-v_i}  
\Psi(\{v_i\})\no\\
&&~~~~~~~+\sum^M_{i=1} \N_i (u,\{v_j\})
\Psi(v_1,\cdots,v_{i-1},u,v_{i+1},\cdots,v_M),\no\\
\eea
with 
\bea
\M_i(u,\{v_j\}) &=& -\frac {\eta}{u-v_i} \a(v_i,M) \prod ^M _{j \neq i} \frac
{v_i-v_j+\eta} {v_i-v_j} , \no\\ 
\N_i(u,\{v_j\}) &=& \frac {\eta}{u-v_i} \delta(v_i,M)\prod ^M _{j \neq i} \frac
{v_i-v_j-\eta} {v_i-v_j} . \no\\ 
\eea
Requiring 
$$\M_i(u,\{v_j\})+\N_i(u,\{v_j\})=0$$ 
forces $\Psi(v_1,\cdots v_M)$ to be an
eigenstate of $\tau(u,0)$ and leads to the Bethe ansatz equations 
\beq
\frac {\a(v_i,M)}{\delta(v_i,M)}=
\prod^M_{j \neq i} \frac {v_i-v_j-\eta}{v_i-v_j+\eta},
~~~~~~i=1,\cdots ,M.
\label {bae-g} \eeq
The corresponding eigenvalue of the matrix $\tau(u,0)$
is
$$
\Lambda (u, \{v_i\}) =
\a (u,M)\prod ^M_{i=1} \frac {u-v_i+\eta}{u-v_i}  
+\delta (u,M)\prod ^M_{i=1} \frac {u-v_i-\eta}{u-v_i} .
$$
\section{Explicit $\Z$-graded realisations}  

Next we give  two nontrivial $\Z$-graded realisations of the algebra
$\A$. One is expressible in terms of two Heisenberg algebras with
generators $a_i,\,a_i^{\d},\,i=1,\,2$ and reads 
$X(u,j)=\tilde{X}(u,j)P_j$ with 
\bea
\tilde{A}(u,j) &=& u^2 + \eta u N +\eta ^2 N_1N_2 -\eta  
(N_1-N_2)\o(N+jI)\no\\
&&-\o^2(N+jI) +a^\d_2 a_1, \no\\
\tilde{B}(u,j) &=& (u+\o(N+jI)+\eta N_1)a_2+\eta^{-1}a_1, \no\\
\tilde{C}(u,j) &=& a^\d_1 (u-\o(N+jI)+\eta N_2)+\eta^{-1}a^{\d}_2, \no \\ 
\tilde{D}(u,j) &=& a_1^{\d}a_2+\eta^{-2}. \label{zg1} \eea 
Above, $P_j$ are the projections defined by (\ref{proj}), 
$N_i=a^{\d}_ia_i$,  $N=N_1+N_2$ and $\o(x)$ is an 
arbitrary polynomial function of $x$. 
Note that in the case when $\o(x)$ is constant, the above realisation
reduces to that discussed in \cite{lz,zlmx} and is factorizable into two local
representations of the Yang-Baxter algebra expressible in terms of the
two Heisenberg algebras. It is
important to note that for generic $\o(x)$ no such factorisation exists. 

The representation acts on the infinite dimensional Fock space
spanned by the vectors 
\beq \left|m,n\right>=(a_1^{\d})^m(a_2^{\d})^n\left|0\right>, ~~~~~~~~
m,\,n=0,1,2,....,\infty . \label{fock} \eeq  
For this representation, we choose the pseudovacuum $|\chi\rangle$ as the Fock
vacuum $|0\rangle$. The representation of the grading operator $Z$ is
chosen to be 
$$\pi(Z)=M.I-N. $$ 
We then have
\beq
\a(u,M)= u^2-\o^2 (M),
~~~\delta(u,M) = \eta ^{-2}
\eeq
and the Bethe ansatz equations  become  
\beq
\eta ^2 (v_i^2-\o^2(M))=
\prod^M_{j \neq i} \frac {v_i-v_j-\eta}{v_i-v_j+\eta}
\eeq
for the diagonalisation of the matrix $\tau(u,0)$. The eigenstates 
(\ref{estates}) in this instance are
also eigenstates of the total particle number $N$ with eigenvalue $M$.

Another $\Z$-graded realisation of the Yang-Baxter algebra 
is $X(u,j)=\tilde{X}(u,j)P_j$ with 
\bea
\tilde{A}(u,j) &=& -\eta u^2 +  
u\left(1-\eta^2 (K_z+N_c) -\eta \o(K_z+N_c+jI)\right)\no\\
&&+\eta K_z -\eta^2 K_z \o(K_z+N_c+jI) -\eta^3 N_cK_z +\eta^2 c K_+, \no\\
\tilde{B}(u,j) &=& \eta (1-\eta u- \eta \o(K_z+N_c+jI)-\eta^2 N_c)K_-
-\eta c(u-\eta K_z), \no\\
\tilde{C}(u,j) &=& \eta c^\dagger (u+\eta K_z)-\eta K_+, \no \\
\tilde{D}(u,j) &=& u-\eta K_z +\eta^2 c^\dagger K_-.\label{su11} 
\eea     
Above, the operators $c,\,c^{\d}$ form a Heisenberg algebra, with
$N_c=c^{\d}c$, and the operators $K_z,\,K_+,\,K_-$ satisfy the relations
of the $su(1,1)$ algebra 
$[K_z,\,K_{\pm}]=\pm K_{\pm},\,\,[K_+,\,K_-]=-2K_z$. 
As in the previous example, $\o(x)$ is an arbitrary polynomial function 
of $x$ and
the above realisation is factorisable only in the case when $\o(x)$ is
constant. 

For this representation, 
we choose the pseudovacuum $|\chi \rangle$ as the tensor
product of the Fock
vacuum $|0\rangle$ with a  lowest weight state for the algebra $su(1,1)$
of weight $\kappa$. The representation of the grading operator may  
be chosen as 
$$\pi(Z)=M.I-K_z-N_c.$$ 
Then, 
\beq
\a(u,M)=(1-\eta u -\eta \o(M))(u+\eta \kappa),
~~~\delta(u,M) = u-\eta \kappa
\eeq
and the Bethe ansatz equations are 
\beq
(1-\eta v_i -\eta \o(M))\left(\frac {v_i+\eta \kappa}{v_i-\eta \kappa}
\right)=
\prod^M_{j \neq i} \frac {v_i-v_j-\eta}{v_i-v_j+\eta}.
\eeq

\section{Three models of Bose-Einstein condensates} 
\subsection{Model 1: Two coupled BECs}

Consider the following general Hamiltonian describing Josephson tunneling
between two coupled Bose-Einstein condensates 
\bea
H&=& U_{11} N_1^2+U_{12}N_1N_2+U_{22} N_2^2 +\mu_1 N_1+\mu_2 N_2\no\\
&& -\frac {\cal {E} _J}{2} (a_1^\dagger a_2 + a_2^\dagger a_1).
\label{jo} \eea
The above Hamiltonian generalises the canonical Josephson Hamiltonian
studied  in
\cite{lz,zlmx} in that the  couplings $U_{11},\,
U_{22}$ for the $S$-wave scattering terms can be chosen arbitrarily.   
It also describes a pair of Cooper pair boxes with
capacitive coupling \cite{mss}. In the limit $U_{22}\rightarrow 0$,
then $\left<N_2\right> >> \left<N_1\right>$, 
which can be considered as a single Cooper
pair box coupled to a reservoir.   

It is an algebraic exercise to show 
that the Hamiltonian is related with the 
matrix $\tilde{\tau}(u,0)=\tilde{A}(u,0)+\tilde{D}(u,0)$ through   
\bea
H&=&
-\frac {\cal {E}_J}{2}
\left [\tilde{\tau} (0, 0) -\eta^{-2} +(\alpha N +\beta)^2
-\eta \sigma N -\eta \delta N^2 \right] 
.\nn \eea
Here we have chosen $\o(N)= \a N+\b$ and 
the coupling constants are
identified as  
\bea
\eta^2&=& \frac{2({U_{11}+U_{22}-U_{12}})}{{\cal {E}_J}}, \no \\
\a &=&\frac{U_{11}-U_{22}}{\eta {\cal {E}_J}},\no \\
\b &=& \frac{\mu_1 -\mu_2}{\eta {\cal {E}_J}}, \no \\
\sigma &=& \frac{\mu_1+\mu_2}{\eta {\cal {E}_J}}, \no \\
\delta&=& \frac{U_{11}+U_{22}}{\eta {\cal {E}_J}} \no \eea 
Noting that 
$$N=\eta ^{-1} \frac{d\tilde{\tau}}{du}(0,0), $$ 
the above demonstrates that the Hamiltonian (\ref{jo}) is expressible
solely in terms of the matrix $\tilde{\tau}(u,0)$ and its derivative. 

Since $[H,\,N]=0$, the
Hamiltonian is block diagonal on the Fock basis (\ref{fock}). Thus on
a subspace of the Fock space with fixed particle number $N$, the
diagonalisation of $\tilde{\tau}(u,0)$ is equivalent to the diagonalisation
of $\tau(u,0)$ presented earlier in the Bethe ansatz framework. We then
deduce that the solution of (\ref{jo}) for the energy spectrum is 
\bea  E&=& 
-\frac{{\cal E}_J}{2}\left[
\eta^{-2}\prod_{i=1}^N\frac{v_i+\eta}{v_i}-\left(\alpha N+\beta   
\right)^2 \prod_{i=1}^N\frac{v_i-\eta}{v_i}   \right. \no \\
&&~~~~~~~~~\left.  -\eta^{-2}+(\a N + \b)^2 
-\eta \sigma N -\eta \delta N^2 
\right]\label{nrg1} \eea 
where the parameters $\{v_i\}$ are subject to the Bethe ansatz equations 
$$\eta^2\left(v_i^2-\left(\alpha N+\beta 
\right)^2\right)=\prod_{j\neq i}^N\frac{v_i-v_j-\eta}{v_i-v_j+\eta}. $$

\subsection{Model 2: Homo-atomic-molecular BECs}    

Next we turn our attention to a two-mode model for 
an atomic-molecular Bose-Einstein condensate with identical
atoms. The Hamiltonian
takes the form
\bea
H&=& U_{aa} N_a^2+U_{ac}N_aN_c+U_{cc} N_c^2 +\mu_a N_a+\mu_c N_c\no\\
&& +\Omega (a^\dagger a^\dagger c + c^\dagger a a) \label{acham}  
\eea
which acts on a basis of Fock states analogous to (\ref{fock}). 
For the case of ${}^{87}$Rb, all of these parameters have been estimated
from experiment (see \cite{cusack}). In the experiment described in
\cite{donley}, the parameter $U_{aa}$ was varied significantly with a
magnetic field.

The Hamiltonian commutes with the total atom number $N=N_a+2N_c$. 
In terms of a  realisation of the algebra $su(1,1)$ through
\beq
K_+ = \frac {(a^\dagger)^2}{2},
\, K_- = \frac {a^2}{2},
\, K_z= \frac {2N_a+1}{4},
\label{acrep} \eeq
one may establish the relation between the Hamiltonian and the
corresponding transfer matrix $\tilde{\tau} (u,0)=\tilde{A}(u,0)+
\tilde{D}(u,0)$ arising from the the realisation (\ref{su11}) of the
Yang-Baxter algebra is 
\bea
H&=&\sigma + \delta (N/2+1/4)+\gamma (N/2+1/4)^2
 + 2\eta ^{-2}\Omega \tilde{\tau} (0,0), \nn \eea
with 
$$\frac{d\tilde{\tau}}{du}(0,0) = 2-\eta (\eta +\a) (N/2+1/4)-\eta \b.$$
Above we have chosen 
\bea \o(K_z+N_c)&=& \a (K_z+N_c) +\b \nn \\
&=&\a (N/2+1/4) +\b \nn \eea  
and the following identification has been made for the coupling constants 
\bea
\eta &=& \frac {4U_{aa}+ U_{cc} -2 U_{ac}}
{2\Omega},\no\\
\a &=& \frac {U_{cc}- 4U_{aa}}
{2\Omega},\no\\
\b &=& \frac {2\mu_c -4\mu_a +4U_{aa}- U_{ac}}
{4\Omega}, \no \\  
\sigma&=& \frac{U_{aa}-2 \mu_a}{4} , \no \\
\delta &=& \frac{2\mu_c-U_{ac}}{2} , \no \\
\gamma &=& U_{cc}  . \no \eea
By the same argument as before, we conclude that the exact solution for
the energy spectrum of (\ref{acham}) is determined by 
\bea
E&=&\sigma +\delta(M+\kappa)+\gamma(M+\kappa)^2 \no \\
&&+ 2\eta ^{-1}\kappa\Omega \left[(1-\eta\left(\alpha (M+\kappa)+\beta)
\right)
\prod_{i=1}^M 
\frac{v_i-\eta}{v_i}-\prod_{i=1}^M\frac{v_i+\eta}{v_i}\right],
\label{acnrg} \eea
where the parameters $v_i$ satisfy the Bethe ansatz equations 
\beq
[1-\eta v_i -\eta (\alpha (M+\kappa)+\beta)]
\left(\frac {v_i+\eta \kappa}{v_i-\eta \kappa}\right)=
\prod^M_{j \neq i} \frac {v_i-v_j-\eta}{v_i-v_j+\eta}.
\label {acbae} \eeq
For the representation (\ref{acrep}) of the $su(1,1)$ algebra there are two
lowest weight vectors; viz. 
the Fock vacuum $\left|0\right>$ and the one particle state $a^\dagger
\left|0\right>$. It follows that the allowed values for $\kappa$ in
(\ref{acnrg},\ref{acbae}) are $\kappa=1/4,\,3/4.$ This demonstrates that the
solution of the model depends on whether the total particle number 
$N=2M+2\kappa-1/2$
is even or odd, the effects of which on the energy spectrum 
can be seen through numerical
analysis (cf. \cite{zlm}). 

\subsection{Model 3: Hetero-atomic-molecular BECs}

The previous construction can be extended to model an 
atomic-molecular Bose-Einstein condensate with two distinct
species of
atoms, denoted $a$ and $b$. For this case the Hamiltonian
takes the form
\bea
H&=& U_{aa} N_a^2+U_{bb} N_b^2 +U_{cc} N_c^2+
U_{ab}N_aN_b+U_{ac} N_aN_c+U_{bc}N_bN_c \no\\
&&+\mu_a N_a+\mu_b N_b+\mu_c N_c
+\Omega (a^\dagger b^\dagger c + c^\dagger b a)
\label{abcham} \eea
which commutes with the total atom number $N=N_a+N_b+2N_c$ and
$\I=N_a-N_b$. 
Here the model acts on the Fock space spanned by the vectors 
$$\left|l,m,n\right>=(a^\dagger)^l(b^\dagger)^m(c^\dagger)^n\left|0\right>.
$$ 
In order to show the solvability of this model, we adopt the 
realisation of the $su(1,1)$ algebra given by 
\beq
K_+ = a^\dagger b^\dagger,
\, K_- = ab,
\, K_z= \frac {N_a+N_b+1}{2},
\eeq
and observe that the operator $\I$ commutes with the $su(1,1)$ algebra
in this representation, hence taking a constant value in any
irreducible representation. Due to the symmetry upon interchanging the
labels $a$ and $b$, we can assume without loss of generality that the
eigenvalues of $\I$ are non-negative. 
In particular, note then that the lowest weight
states for this realisation are of the form 
$$\left|m\right>=(a^\dagger)^m\left|0\right>,~~~~~m=0,1,2,.....$$ 
and $K_z\left|m\right>=(m/2+1/2)\left|m\right>$.  
We conclude that the lowest weight labels $\kappa$ can be taken from
the set $\{1/2,\,1,\,3/2,....\}$ and the eigenvalue of $\I$ on
the irreducible representation labelled by $\kappa$ is $2\kappa-1$.

For this case the relation between the Hamiltonian and the
corresponding matrix $\tilde{\tau} (u,0)$ is   
\bea
H&=&
\sigma+\delta(N/2+1/2)+\lambda(N/2+1/2)^2 \no \\
&&~~+ \rho \I + \nu \I^2 +
\xi \I (N/2+1/2) 
+ \eta ^{-2}\Omega \tilde{\tau} (0,0)
\eea
with 
$$\frac{d\tilde{\tau}}{du}(0,0) = 2-\eta (\eta +\a) (N/2+1/2)-\eta \b
\I-\eta \g.$$
Above we have chosen 
\bea \o(K_z+N_b)&=& \a (K_z+N_c) +\b (2\kappa -1)+\g\nn \\ 
&=&\a (N/2+1/2) +\b \I +\g\nn \eea 
and 
the coupling constants are related through the relations   
\bea
\eta &=& \frac{U_{aa}+U_{bb}+U_{cc}+U_{ab}-U_{ac}-U_{bc}}{\Omega}  , \nn \\
\a &=&\frac{U_{cc}-U_{aa}-U_{bb}-U_{ab}}{\Omega}  , \nn \\
\beta &=& \frac{2U_{bb}-2U_{aa}+U_{ac}-U_{bc}}{2\Omega}, \no \\
\gamma &=&\frac{2U_{aa}+2U_{bb}+2U_{ab}-U_{ac}-U_{bc}+2\mu_c-2\mu_a-2\mu_b}
{2\Omega}  , \nn \\
\sigma &=&\frac{U_{aa}+U_{bb}+U_{ab}-2\mu_a-2\mu_b}{4} , \nn \\
\delta &=&\frac{2\mu_c-U_{ac}-U_{bc}}{2}   , \nn \\ 
\lambda &=&U_{cc}  , \nn \\
\rho &=&\frac{U_{bb}-U_{aa}+\mu_a-\mu_b}{2}  , \nn \\
\nu &=&\frac{U_{aa}+U_{bb}-U_{ab}}{4}  , \nn \\
\xi &=&\frac{U_{ac}-U_{bc}}{2}   
.\no
\eea
The exact solution in this instance reads 
\bea
E&=&
\sigma +\delta(M+\kappa)+\lambda(M+\kappa)^2\no \\
&&~~+\rho(2\kappa-1)+
\nu(2\kappa-1)^2+\xi(2\kappa-1)(M+\kappa) \no \\
&& ~~+ \eta ^{-1}\kappa\Omega 
\left[(1-\eta(\alpha (M+\kappa)+\beta(2\kappa-1)+\gamma))
\prod_{i=1}^M
\frac{v_i-\eta}{v_i}\right. \no \\ 
&&~~~~~~~~~~~~~~~~~~~~~~~~~~~~~~~
~~\left.-\prod_{i=1}^M\frac{v_i+\eta}{v_i}\right],
\nn \eea
where the parameters $v_i$ satisfy the Bethe ansatz equations
\beq
[1-\eta v_i -\eta \left(\alpha (M+\kappa)+\beta(2\kappa-1)+\gamma\right)]
\left( \frac {v_i+\eta \kappa}{v_i-\eta \kappa}\right)=
\prod^M_{j \neq i} \frac {v_i-v_j-\eta}{v_i-v_j+\eta}
\label {abcbae} \eeq
and the total atom number is given by $N=2M+2\kappa -1$. 

\section{Wave function scalar products} 

Recall that in the usual algebraic Bethe ansatz for the algebra $\A$
there is a formula originally due to Slavnov \cite{slavnov} (see also
\cite{kib,kmt}) for the wave function scalar products. 
The Slavnov formula still applies in the $\Z$-graded case and takes
the usual form 
\bea
S_M(\{u_j\},\{v_k\}) &=&
\Phi(\{u_j\})\Psi(\{v_k\})  \no\\
&=&
\Phi(\{v_k\})\Psi(\{u_j\})  \no\\
&=&\frac {{\rm det} T(\{u_j\},\{v_k\})}{{\rm det} V(\{u_j\},\{v_k\})},
\eea
with the entries of the $M\times M$ matrices $T$ and $V$ given by 
$$
T_{ab}= \frac {\partial}{\partial v_a} \Lambda (u_b,\{v_k\}),
V_{ab} = \frac {1}{u_b-v_a}, a,b=1,\cdots,M
$$
$\Phi(\{u_i\})$ is the left vector 
$$\Phi(u_1,\cdots,u_M)=\left<\chi\right|B(u_M,M)\cdots B(u_1,1). $$
Above, we have adopted the usual convention to scale the Yang-Baxter
algebra such that $\delta(u,M)=1$. 
Also,  $\{ v_k \}$ provide  a solution of the Bethe ansatz
equation (\ref{bae-g}) and the parameters $\{ u_j \}$ can be chosen
arbitrarily.

The Yang-Baxter algebra $\A$ admits a conjugation operation $\dagger:\A
\rightarrow \A$   defined by 
$$A(u)^\dagger=A(u),~~~B(u)^\dagger=C(u),~~~C(u)^\dagger=B(u),~~~
D(u)^\dagger=D(u)$$ 
and extended to all of $\A$ through 
$$(\theta.\phi)^\dagger=\phi^\dagger.\theta^\dagger, ~~~
\forall\, \theta,\,\phi\,\in \A$$
such that the defining relations (\ref{YBA}) are preserved.  
Consequently the right vector $\Phi(v_1,\cdots,v_M)^\dagger$ 
is also an eigenvector of the transfer matrix whenever the Bethe ansatz
equations for the parameters $\{v_i\}$ are satisfied. 
However, it is apparent that the $\Z$-graded representations 
(\ref{zg1},\ref{su11}) 
we have introduced are not unitary, and generally  
$$\Phi(\{v_i\})^\dagger\neq \Psi(\{v_i\}). $$ 
On the other hand, numerical analysis we have undertaken for the above
models indicates that for fixed particle numbers, and 
generic values of the coupling parameters, the
energy spectrum is free of degeneracies. This is presumably due to the
fact that the only Lie algebra symmetries for these models are $u(1)$
invariances corresponding to conservation of particle numbers, and the
non-degenerate spectra are examples of Hund's non-crossing rule 
\cite{hund,vnw}. 
Whenever this is the case, we can conclude that 
$$\Phi(\{v_i\})^\dagger = K  \Psi(\{v_i\}) $$
for some constant $K$ and the Slavnov formula can still be invoked for
the computation of form factors and correlation functions 
(cf. the example of \cite{lz}
where it was found $K=\pm 1$). 

\section{Conclusion}

In conclusion we have introduced a new scheme for the algebraic Bethe ansatz
to diagonalise three classes of 
integrable models relevant to Bose-Einstein condensates of dilute alkali
gases. The extension of this construction to other types of models,
such as the Jaynes-Cummings model \cite{jm}, and generalised 
Tavis-Cummings
model discussed in \cite{rktb},  
is straightforward.
~~\\
~~\\
~~\\
%\vskip.3in
%\acknowledgments
This work is supported by  the Australian Research Council.
%\newpage
%\vskip.3in

\end{document}